\documentclass[doublecol]{epl2}
\usepackage{times}
\usepackage{color}
\usepackage{graphicx}
\usepackage{dcolumn}
\usepackage{amssymb}
\usepackage{amsmath}
\usepackage{amsfonts}
\usepackage{bm}%
\usepackage[leftcaption]{sidecap}
\usepackage{wrapfig}
\usepackage[colorlinks=true,linkcolor=blue,pagecolor=blue,filecolor=blue,menucolor=blue,urlcolor=blue,citecolor=blue,anchorcolor=blue]{hyperref}%

\title{Strain Controlled Spin and Charge Pumping in Graphene Devices via Spin-orbit Coupled Barriers}

\author{Ramin Mohammadkhani\inst{1}, Babak Abdollahipour\inst{2} \and Mohammad Alidoust\inst{3,4} }

\institute{
  \inst{1} Department of Physics, Faculty of Science,
           University of Zanjan, Zanjan 45371-38791, Iran\\
\inst{2} Faculty of Physics,
          University of Tabriz, Tabriz 51666-16471, Iran\\
\inst{3} Department of Physics, University
        of Basel, Klingelbergstrasse 82, CH-4056 Basel, Switzerland\\
\inst{4} Department of Physics,
        Faculty of Sciences, University of Isfahan, Hezar Jerib Avenue,
        Isfahan 81746-73441, Iran}


\abstract{ We theoretically propose a graphene-based adiabatic
quantum pump with intrinsic spin-orbit coupling (SOC) subject to
strain where two time-dependent extrinsic spin-orbit coupled
barriers drive spin and charge currents. We study three differing
operation modes where $i$) location, $ii$) chemical potential, and
$iii$) SOC of the two barriers oscillate periodically and out of
phase around their equilibrium states. Our results demonstrate that
the amplitude of adiabatically pumped currents highly depends on the
considered operation mode. We find that such a device operates with
highest efficiency and in a broader range of parameters where the
barriers' chemical potential drives the quantum pump. Our results
also reveal that by introducing strain to the system, one can
suppress or enhance the charge and spin currents separately,
depending on strain direction.}

\begin{document}

\maketitle Spintronics is an emerging filed which has aimed at
exploiting the spin degree of freedom to construct faster and high
performance low-power nanoscale devices\cite{cite:Zutic04}. The
discovery of isolated graphene
monolayer\cite{cite:Novoselov04,cite:Zhang05}, a single layer of
Carbon atoms, with unique electrical, optical and thermal properties
has triggered numerous efforts to achieve graphene-based nanoscale
devices\cite{cite:Han14,cite:Avsar14,RSOC1,RSOC2}. The massless
Dirac fermions in ballistic graphene can reflect chirality and
linear dispersion relation of graphene around the Dirac points; two
inequivalent corners of the first Brillouin zone\cite{cite:Neto09}.
Also, the long spin relaxation time of the Dirac fermions in
graphene monolayers due to a small intrinsic spin-orbit coupling
(SOC) which originates from the intra-atomic spin-orbit coupling of
the Carbon atoms has made it an exceptional candidate to the
spintronics devices\cite{cite:Han14}.

Quite recently, it was experimentally demonstrated that a strong Rashba SOC $\sim 17$ meV can be induced into
graphene monolayers by means of proximity to a semiconducting tungsten disulphide
substrate\cite{cite:Avsar14}. This finding is highly appealing in terms of generation and manipulation of spin
currents in more controllable platforms. The intrinsic SOC that can be caused by the crystalline potential
associated with the band structure respects all the lattice symmetries in graphene and results in a small
energy gap at the Dirac points. The extrinsic or Rashba SOC, however, results from the lack of inversion
symmetry due to perpendicular electric fields, substrate effects, chemical doping, or curvature of graphene
corrugations and can be responsible for inducing a spin polarization in graphene.\cite{Kane05,so_exp1} The
influences of intrinsic and Rashba SOCs on the transport properties of graphene monolayer systems have
extensively been studied in the recent years\cite{cite:Dyrdal,cite:Bercioux,cite:Gurjic,RSOC1,RSOC2,RSOC4}.
For instance, it was shown that spin polarization induced by a charge current can reside in the graphene plane
and perpendicular to the electric field while its sign changes by varying the Fermi level through an external
gate voltage\cite{cite:Dyrdal}. Also, it was theoretically found that the interplay of massive electrons with
SOC or strain in a graphene layer can result in a spin-valley filter\cite{cite:Gurjic,Zhai}.

Spin and charge quantum pumpings are striking topics in the context of quantum transport through
nanostructures. The quantum nature of these effects arises from the geometric (Berry) phases and quantum
interference effects \cite{cite:Makhlin01}. An adiabatically pumped current requires, at least, two parameters
of system vary periodically and out of phase in time \cite{cite:Brouwer98}. The adiabaticity is achieved when
the characteristic time of the variations is much smaller than the dwell time of carriers. In this base,
several proposals for charge pumping through graphene junctions were introduced during the past
years\cite{cite:Pradaprb09,cite:Grichuk13,cite:Zhu09,cite:Low12,
cite:abdollahipour14,Benjamin,pmp1,pmp2,pmp3}.

Motivated by the recent researches on time-dependent graphene
systems\cite{cite:Pradaprb09,cite:Grichuk13,cite:Zhu09,cite:Low12,
cite:abdollahipour14,Benjamin,pmp1,pmp2,pmp3} and experimentally achieved graphene layers with strong
extrinsic spin-orbit couplings\cite{cite:Avsar14,so_exp1,so_exp2}, in this paper we propose a novel device to
generate controllable charge and spin pumped currents without resorting to any externally imposed field. This
device consists of a graphene monolayer with length $2L$ and width $W$ under strain with intrinsic spin-orbit
coupling and the pumped currents are driven by two extrinsic spin-orbit coupled barriers induced by a
substrate\cite{cite:Avsar14}.

We assume that the chemical potential/ location or SOC of the barriers can be time-dependent and periodically
oscillate out of phase. Our results reveal that the quantum pump operates with highest efficiency where the
barriers chemical potential drives the currents. It is shown that, in the latter case, the currents' amplitude
is more pronounced and the quantum pump operates in a broader range of the system parameters. We also uncover
how an in-plane strain in the graphene layer alters and controls the spin and change currents simultaneously.
Our results demonstrate that a weak strain applied to the graphene plane can enhance the spin current and
suppress the charge current simultaneously, depending on the direction of strain.

\begin{figure}
\centerline{\includegraphics[width=8.cm]{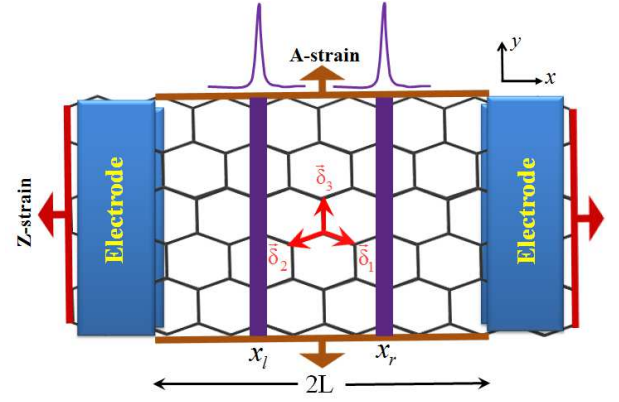}} \caption{\label{fig:model} Schematic of the strained
graphene quantum pump proposed in this paper. Two time-dependent barriers located at $x_l$ and $x_r$ pump spin
and charge currents across the system which are modeled by the Dirac delta function. The system contains $5$
regions ($\gamma$) from left to right: $(\gamma)=(1),(5)$ are electrode regions and $(\gamma)=(2),(3),(4)$
regions are the ISO coupled graphene segments separated by the vibrating barriers. We assume that the strain
imposed can be either in $A$- or $Z$-direction and label them by $A$-strain/$Z$-strain.}
\end{figure}

The quasiparticles at low energies in a monolayer of graphene under
tension and in the presence of intrinsic and extrinsic SOCs (ISO and
ESO) are governed by the following Hamiltonian
\cite{cite:Bercioux,Kane05}:
\begin{eqnarray}\label{eq:hamil1}
&&{\cal H}= {\cal H}_0 + {\cal H}_{\text{ISO}}+{\cal
H}_{\text{ESO}};\\
&&{\cal H}_0= v_xp_x s_0\sigma_x
+v_yp_y s_0\sigma_y+{
\mu}(x) s_0\sigma_0,\nonumber\\
&&{\cal H}_{\text{ISO}}=\beta s_z\sigma_z,\;\;\; {\cal
H}_{\text{ESO}}=\alpha[s_y\sigma_x-s_x\sigma_y].\nonumber
\end{eqnarray}
Here, $\mu(x)$ is a tuneable chemical potential which can be
controlled by an external gate voltage. $\beta$ and $\alpha$ are the
strength of intrinsic and extrinsic SOCs, respectively. $\sigma_0$
and $s_0$ denote $2\times 2$ unitary matrices, $\sigma_i$ and $s_i
(i={x,y,z})$ are the Pauli matrices in the pseudospin and the real
spin subspaces, respectively. The proposed quantum pump is
schematically depicted in Fig. \ref{fig:model}. There are two
electrode regions where carriers' density can be externally
controlled. The entire of graphene layer is assumed intrinsically
spin-orbit coupled. We focus on strains applied in two distinct
crystallographic directions: zig-zag ($Z$) and armchair ($A$) as
shown in Fig. \ref{fig:model}. In order to study the influences of
strain on the characteristics of system transport, we adopt a model
introduced in Ref. \cite{cite:Pereira} and expand tight-binding
result for the band structure with arbitrary hopping energies
$t_{1,2,3}$ around the new Dirac point
$\mathbf{K}_D=(\cos^{-1}(-1/2\eta)/\sqrt{3}a_x,0)$, namely,
$E_k=\pm| \sum_{i=1}^{3}t_ie^{-i\vec{k}\cdot\vec{\delta}_{i}}|$
\cite{cite:choi1,cite:soodchomshom}. As shown in Fig.
\ref{fig:model}, $\vec{\delta}_{i}$ are displacement vectors between
two nearest neighbor Carbon atoms. We assume
$t_{1}=t_{2}=\tilde{\epsilon}$ and $t_3=\epsilon$ in our
calculations and set $\eta$ equal to ratio
$\tilde{\epsilon}/\epsilon$. The quasiparticles' velocities are
given by $v_x=2\tilde{\epsilon}a_x\sin(\cos^{-1}(-1/2\eta))/\hbar$
and $v_y=3\epsilon a_y/2\hbar$ \cite{cite:choi1,cite:soodchomshom}.
The hoping energies are given by
$t_{i}=\epsilon_{0}e^{-3.37(\mid\vec{\delta}_{i}\mid)/c_{0}-1}$, where $c_0=0.142 {\AA}$ and $\epsilon_{0}$
are the distance of two Carbon atoms and hoping energy in the undeformed graphene layer, respectively. The
displacement vectors under the $Z$-strain leads to $\vec{\delta}_1=a_{x}\sqrt{3}\hat{x}-a_{y}\hat{y}$,
$\vec{\delta}_2=-a_{x}\sqrt{3}\hat{x}-a_{y}\hat{y}$, $\vec{\delta}_3=2a_{y}\hat{y}$ while for the armchair
($A$-strain), $\vec{\delta}_1=a_{y}\sqrt{3}\hat{x}-a_{x}\hat{y}$,
$\vec{\delta}_2=-a_{y}\sqrt{3}\hat{x}-a_{x}\hat{y}$, $\vec{\delta}_3=2a_{x}\hat{y}$. Here $a_{x}=(1+ s)c_0 /2$
and $a_{y}=c_0(1- ps)/2$ with $p=0.165$ which is the Poisson's ratio for graphite and $s$ represents the
strength of applied tension. The {\it maximum} tension strength considered throughout this paper can be less
than the gap threshold value, i.e. $s\leq0.23\sim 20\%$ predicted theoretically (see
Ref.\cite{cite:Pereira,liuprb2007}). The experimental evidence for maximum strain exerted on graphene without
change in its band structure is less than $\sim 15\%$\cite{Lee08}. Nonetheless, we emphasize that a maximum of
$\sim 15\%$ does not affect the main conclusions of this paper. We have used the generalized Weyl-Hamiltonian
which is in a very good agreements with the \textit{ab initio} calculations. If we diagonalize the total Dirac
Hamiltonian ${\cal H}$ given by Eq. (\ref{eq:hamil1}), we arrive at the following spinors in each region of
the system represented by $\gamma=1,2,3,4,5$ (see Fig. \ref{fig:model})
\begin{eqnarray}
&&\psi_{+}^{\pm}(x_\gamma,\varepsilon_n)=(1,\pm\zeta_{n,+}^{\gamma,\pm},0,0)
e^{i(\pm\kappa_{n}^{\gamma}x_\gamma+q_ny)}
\\&&
\psi_{-}^{\pm}(x_\gamma,\varepsilon_n)=(0,0,1,\pm\zeta_{n,-}^{\gamma,\pm})
e^{i(\pm\kappa_{n}^{\gamma}x_\gamma+q_ny)}\\
&&\kappa^\gamma_n=\Big(\frac{\varepsilon_{F\gamma}^2-\beta_\gamma^2-\hbar^2
v_{y\gamma}^2q_n^2}{\hbar^2 v_{x\gamma}^2}\Big)^{1/2},\;
\zeta_{n,\sigma}^{\gamma,\pm}=\frac{v_{x\gamma}\kappa^\gamma_n\pm i v_{y\gamma}q_n}
{\sigma\beta_\gamma+\varepsilon_{F\gamma}}
\nonumber
\end{eqnarray}
where $\sigma=\pm$, $\varepsilon_{F\gamma}$ is the quasiparticles' energy measured from the chemical potential
level, and $q_n$ stands for the transverse component of the wave vector which is conserved in different
$\gamma$ regions. The junction is assumed sufficiently wide, $W\gg L$, which allows for replacing $\sum_{q_n}$
by $\int dq$ in our calculations. For numerical purposes, we define the total wave vector ${\cal
    K}_n^\gamma$, i.e. $\kappa_{n}^{\gamma}={\cal
    K}_n^\gamma\cos\theta_{n,\gamma}$ and $q_{n}={\cal
    K}_n^\gamma\sin\theta_{n,\gamma}$, as follows:
\begin{subequations}
\begin{eqnarray}
    &&\hbar{\cal
    K}_n^\gamma=\Big(\frac{\varepsilon_{F\gamma}^2-\beta_\gamma^2}{v_{x\gamma}^2
    \cos^2\theta_{n,\gamma}+v_{y\gamma}^2\sin^2\theta_{n,\gamma}}\Big)^{1/2},\\&&
\zeta_{n,\sigma}^{\gamma,\pm}=
\sqrt{\frac{\varepsilon_{F\gamma}-\sigma\beta_{\gamma}}
{\varepsilon_{F\gamma}+\sigma\beta_{\gamma}}}\frac{v_{x\gamma}\cos\theta_{n,\gamma}\pm
iv_{y\gamma}\sin\theta_{n,\gamma}}
{\sqrt{v_{x\gamma}^2\cos^2\theta_{n,\gamma}+v_{y\gamma}^2\sin^2\theta_{n,\gamma}}},\nonumber\\&&\\&&
\theta_{n,\gamma}=\arcsin\Big[\frac{\hbar^2
q_n^2v_{x\gamma}^2}{\varepsilon_{F\gamma}^2-\beta_\gamma^2+(v_{x\gamma}^2-v_{y\gamma}^2)q_n^2}\Big]^{1/2}.
\end{eqnarray}
\end{subequations}
To model the vibrating barriers shown in Fig. \ref{fig:model}, we assume that experimentally tuneable
parameters at the barriers are $i$) the chemical potentials $\mu_{l,r}$ and $ii$) the ESOC $\alpha_{l,r}$ in
addition to $iii$) their locations $x_{l,r}$. The first mode, $i$, of the pumping can be realized by tuning
the potential of underlying gats. To experimentally realize the two other modes one may construct the barriers
through two flexible cantilevers with vertical and horizontal oscillations around their equilibrium locations
on top of the graphene sheet, respectively. The total wave function of a particle $\Psi$ passing through the
barriers experiences the following transformation ${\cal T}_{l,r}$ at the left ($l$) and right ($r$) barriers
namely, $\Psi^{\cal R}= {\cal T}_{l,r} \Psi^{\cal L}$ in which
\begin{eqnarray}\label{eq:trnsfm}
    {\cal T}_{l,r}=\frac{2 i\hbar v_x^{\cal L}
    s_0\sigma_x+\mu_{l,r}s_0\sigma_0+
    \alpha_{l,r}(s_y\sigma_x-s_x\sigma_y)}
    {2 i\hbar v_x^{\cal    R}s_0\sigma_x-\mu_{l,r}
     s_0\sigma_0-\alpha_{l,r}(s_y\sigma_x-s_x\sigma_y)}.
\end{eqnarray}
The transformations ${\cal T}_{l,r}$ are derived by integrating the
Dirac Hamiltonian Eq. (\ref{eq:hamil1}) over the $x$-direction in close vicinities
of the barriers and modeling the vibrating barriers through spatial Dirac
deltas.\cite{cite:Titov07,cite:abdollahipour14}. $v_x^{\cal L}$
and $v_x^{\cal R}$ show the velocity of particles at the left
(${\cal L}$) and right (${\cal R}$) sides of the barriers. The
charge and spin currents pumped by ${\cal X}_{l,r}$: two
periodic and out of phase oscillating parameters at the barriers (and
within the bilinear response regime where $\delta {\cal X}_{l,r}\ll
{\cal X}_{l,r}$) can be expressed
by\cite{cite:Brouwer98,cite:abdollahipour14}:
\begin{eqnarray}
  &&I^{\cal X}_{\text{charge}}(I^{\cal X}_{\text{spin}})=
  \nonumber\\&&N_mI_0\sum_{\sigma=\pm}\int_{-\infty}^{+\infty}\frac{dq}{{\cal
  K}_F}(\sigma) \text{Im}\Big\{  \frac{\partial {\frak R}_{\sigma}}{\partial{{\cal X}_{l}}}
  \frac{\partial {\frak R}_{\sigma}^\ast}{\partial{{\cal
  X}_{r}}}+  \frac{\partial {\frak T}_{\sigma}}{\partial{{\cal X}_{l}}}
  \nonumber\frac{\partial {\frak T}_{\sigma}^\ast}{\partial{{\cal X}_{r}}}\Big\},\\
\end{eqnarray}
in which the pumping parameters oscillate around equilibrium values ${\cal X}_{l,r}(0)$ and are given by
${\cal X}_{l,r}(t)={\cal X}_{l,r}(0)+\delta{\cal X}_{l,r}\cos(\Omega t+\varphi_{l,r})$. To reside in the
adiabatic regime, the pumping frequency should be of terahertz range, i.e., $\Omega/2\pi\sim$1 THz
\cite{cite:Pradaprb09}. The spin-dependent reflection and transition coefficients are denoted by ${\frak
R}_{\sigma}$ and ${\frak T}_{\sigma}$, respectively. Here $I_0=0.5\pi^{-1}\Omega e \delta{\cal
X}_{l}\delta{\cal X}_{r}\sin\varphi$ in which $\varphi=\varphi_{r}-\varphi_{l}$ is the phase difference of two
oscillating parameters. In what follows, we normalize the currents by $I_0N_m$, and thus define $I^{\cal
X}_c=I^{\cal X}_\text{charge}/I_0N_m$, $I^{\cal X}_s=I^{\cal X}_\text{spin}/I_0N_m$ where $N_m$ is the number
of available modes at the fermi level.
\begin{figure}
\includegraphics[width=8.15cm]{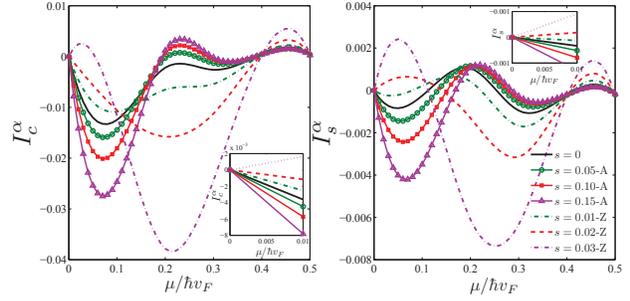}
\caption{\label{fig:I_da} Adiabatically pumped charge and spin
currents: $I^{\alpha}_c$ and $I^{\alpha}_s$ as a function of
chemical potential at the left and right barriers $\mu_l=\mu_r=\mu$.
The time dependent parameters are the strength of ESOCs at the left
and right barriers ($\alpha_{l,r}$) and their equilibrium strengths
are set fixed at $\alpha_l=\alpha_r=3.5\hbar v_F$. Two kinds of
strain ($A$ and $Z$) are considered at three differing values of
strain $s$: $A$-strain, $s=0.05, 0.10, 0.15$ and $Z$-strain,
$s=0.01, 0.02, 0.03$.}
\end{figure}

Figure \ref{fig:I_da} exhibits the charge and spin currents
adiabatically pumped where the strength of ESOCs at the right and
left barriers serve as the pumping parameters with the same
equilibrium values $\alpha_l=\alpha_r=\alpha$ and
$\mu_{l}=\mu_{r}=\mu$. The parameters of the barriers are set at
$\alpha_{l,r}=3.5\hbar v_F$, $x_{l}=0.0$, $x_{r}=0.5L$, while the
chemical potential of regions $\gamma=2,3,4$ are considered fixed at
${\mu}_{2,3,4}=0.5\hbar v_F$. The intrinsic spin-orbit coupling is
assumed constant throughout the graphene layer $\beta=0.02\hbar v_F$
which is equivalent to $\approx 0.05$meV and the extrinsic SO is
about $\alpha \approx 9.0$meV\cite{so_exp1,so_exp2}
. The pumped currents are plotted against the doping level of the
barriers i.e. $\mu_{r,l}$ normalized by $\hbar v_F$. Here $2L$ is
the junction length (see Fig. \ref{fig:model}) and $v_F$ is the
velocity of Dirac fermions at the fermi level in an undeformed
graphene sheet i.e. $v_x=v_y=v_F$. The left and right panels show
the charge and spin currents ($I_\text{charge}^{\alpha}$ and
$I_\text{spin}^{\alpha}$ normalized by $N_mI_0$), respectively. The
inset panels are close-ups of the currents where the barriers'
chemical potential is restricted to $0<\mu<0.01\hbar v_F$. The solid
black lines exhibit the currents where no strain is exerted to the
system ($s=0$), in contrast to the other curves which show the
effect of the in plane strain imposed to the graphene layer. We have
considered both armchair and zig-zag strains as sketched in Fig.
\ref{fig:model}. To have similar magnitudes for the pumped currents,
we set $s=0.05,0.10,0.15$ for the strength of the $A$-strain while
$s=0.01,0.02,0.03$ for the $Z$-strain. The values of $s$ considered
here ensure that the strain is enough weak $s<20\%$, so that no gap
opens in the particles' energy
spectrum\cite{cite:choi1,aldst_prb_strain,cite:soodchomshom,Covaci}.
As seen, the pumped charge current is one order of magnitude greater
than the spin current. Figure \ref{fig:I_da} illustrates how the
charge and spin currents can be manipulated through applying strain
in different directions to the device. The spin current can change
sign while the charge current direction remains intact upon moving
from $s=0.01$-$Z$ to $0.02$-$Z$ at $\mu\sim 0.05\hbar v_F$. By
increasing the tension the overall amplitudes of the pumped currents
for both $A$- and $Z$-directions enhance.  Also, the inset panels
reveal that the charge and spin currents are zero at $\mu=0$
independent of the strain direction applied.

\begin{figure}
\centerline{\includegraphics[width=8.15cm]{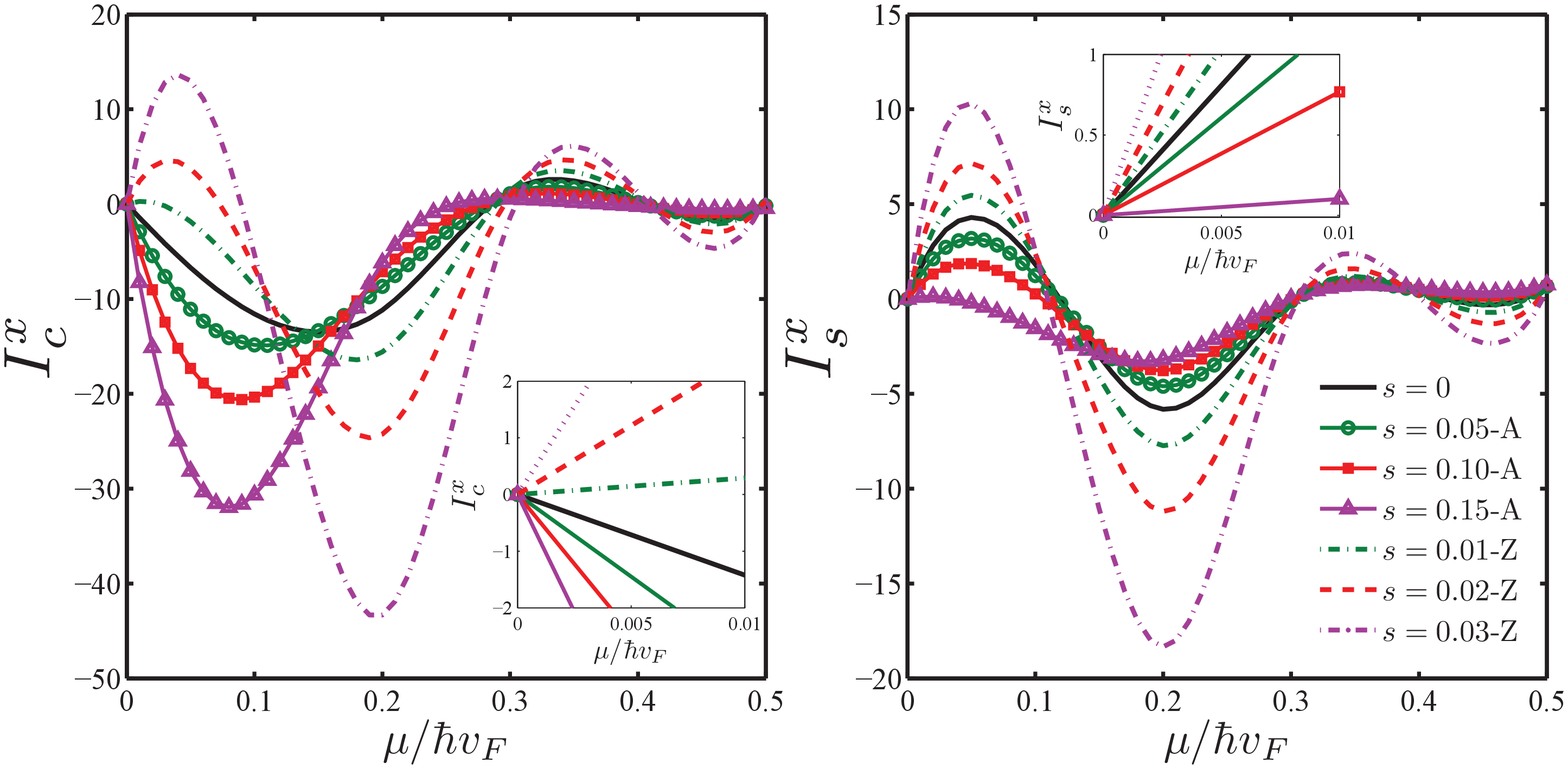}}
\caption{\label{fig:I_dx} Adiabatic spin and charge pumping vs
$\mu_l=\mu_r=\mu$ where the location of barriers $x_{l,r}$ vibrate
around their equilibrium values. The equilibrium location of
barriers are set at $x_l=0$, $x_r=0.5L$ and the strength of ESCOs
fixed at $\alpha_l=\alpha_r=3.5\hbar v_F$. $s=0.05, 0.10, 0.15$ and
$s=0.01, 0.02, 0.03$ are values considered for the strength of $A$-
and $Z$ strains applied to the device, respectively.}
\end{figure}
Figure \ref{fig:I_dx} shows the adiabatically pumped charge and spin
currents where the location of the barriers vibrate out of phase in
time: $I^{x}_c$ and $I^{x}_s$, vs the normalized barriers' chemical
potential $\mu_{l}=\mu_{r}=\mu$.
The parameters are set identical to those of Fig. \ref{fig:I_da}.
Here, the pumped charge and spin currents have the same order of
magnitudes, but they are at least three orders of magnitude greater
than the pumping through time dependent ESOCs at the barriers (see
Figs. \ref{fig:I_da} and \ref{fig:I_dx}). Unlike the pumped currents
generated by oscillating ESOCs where strain has similar effects on
the spin and charge currents, the increment of strain strength in
the $A$-direction here induces an overall enhancement in the pumped
charge current while it causes an overall suppression in the spin
current. However, the increment of strain strength in $Z$-direction
has similar effects on the pumped spin and charge currents and
causes overall enhancement in both the spin and charge currents.
Similar to Fig. \ref{fig:I_da}, we see that the charge and spin
currents are zero at $\mu_l=\mu_r=0$ which is clearly apparent in
the inset panels.
\begin{figure}
\centerline{\includegraphics[width=8.15cm]{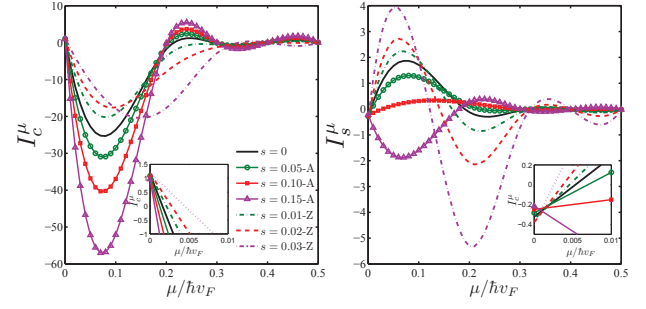}}
\caption{\label{fig:I_du} Adiabatically pumped charge and spin
currents where the chemical potentials $\mu_{l,r}$ at the barriers
vibrate around their equilibrium values: $I^{\mu}_c$ and
$I^{\mu}_s$. The locations and ESOCs of the barriers are assumed
fixed at $x_l=0$, $x_r=0.5L$, and $\alpha_l=\alpha_r=3.5\hbar v_F$,
respectively. The strength of strain applied to the device is equal to
$s=0.05, 0.10, 0.15$ in the $A$-strain and $s=0.01, 0.02, 0.03$ in
the $Z$-strain classes.}
\end{figure}
We now turn to the most important operation mode, namely where the pumping parameters are the chemical
potentials at the barriers $\mu_{l,r}$, oscillating adiabatically around their equilibrium values. Results are
shown in Fig. \ref{fig:I_du} and all of the parameters are set at identical values to the two previous cases.
We find that the amplitudes of the pumped currents have the same order of magnitudes as Fig. \ref{fig:I_dx}
where the location of barriers vibrate around their equilibrium values. As in the previous cases, strain has
pronounce influences on the pumped charge and spin currents. Further investigations demonstrate that
increasing the strain strength in the $A$-direction increases the overall amplitude of the pumped charge
current, but decreases the overall amplitude of the spin current. On the other hand, the increment of the
$Z$-strain has opposite effects on the spin and charge currents. These unequal effects of the strain on the
pumped charge and spin currents offers an experimentally feasible fashion to manipulate and control pumping of
the charge and spin currents through the device. The inset panels illustrate the behavior of charge and spin
currents near $\mu_{l,r}\sim 0$. In contrast to the previous cases, we see that the charge and spin currents
are nonzero at $\mu_{l,r}=0$. The absence of threshold value in chemical potential to generate the currents
and nonzero value of the current pumped in Fig. \ref{fig:I_du} result from the linear dispersion relation and
chiral nature of Fermions in grphene\cite{cite:Pradaprb09,cite:abdollahipour14}. In effect, our further
investigations demonstrate that to generate and manipulate the spin current, the chemical potential should be
`{\it nonzero}'. This is shown by the inset panels of Figs. \ref{fig:I_da} and \ref{fig:I_dx} where no spin
current passes through the system at the zero chemical potential $\mu=0$. This issue however disappears where
the chemical potentials oscillate around an equilibrium value even at $\mu=0$. Hence, by considering the
amplitudes of adiabatically pumped currents ($I_{c,s}^{\alpha}$, $I_{c,s}^{x}$, $I_{c,s}^{\mu}$) and the
manipulation of spin currents over a wide range of $\mu_{l,r}$, one concludes that oscillating chemical
potentials at the barriers would provide more effective mechanism to generate spin current and control the
magnitude of charge and spin currents through strain.

In conclusion, we have proposed a novel quantum pump consisting of a
strained graphene monolayer with intrinsic spin-orbit coupling and
two vibrating extrinsic spin-orbit coupled barriers. To generate
adiabatically pumped currents we consider three different operation
modes to the device: $i$) the strength of extrinsic spin-orbit
couplings, $ii$) the locations and $ii$) the chemical potential of
the barriers oscillate out of phase in time. We have shown that such
a device operates with largest amplitude of pumped currents where
the chemical potential of the barriers oscillates in time and drives
the charge and spin currents into the system. Our results have found
that this operation mode has also broader functionality range in
terms of parameters compared with the other modes. Our study
revealed that a strain applied to the plane of graphene layer can
play key roles to control and manipulate the spin and charge
currents separately, namely one can tune the spin current and
suppress the charge current simultaneously. This interesting
phenomenon originates from the opposite effects of strain on the
pumped spin and charge currents, depending on the direction of
strain imposed to the device.


\begin{thebibliography}{0}
\bibitem{cite:Zutic04}
I. \v{Z}uti\'{c}, J. Fabian and S. D. Sarma, Rev. Mod. Phys. {\bf 76}, 323 (2004).

\bibitem{cite:Novoselov04}
S. K. Novoselov, et. al., Science {\bf 306}, 666 (2004).

\bibitem{cite:Zhang05}
Y. Zhang, W. Y. Tan, L. H. Stormer and P. Kim, Nature, {\bf 438},
201 (2005).

\bibitem{cite:Neto09}
H. A. Castro Neto, F. Guinea, R. M. N. Peres, S. K. Novoselov and K. A. Geim,
Rev. Mod. Phys., {\bf 81}, 109 (2009).

\bibitem{RSOC1} Kh. Shakouri, M. Ramezani Masir, A. Jellal, E. B. Choubabi,
and F. M. Peeters, Phys. Rev. B  {\bf 88}, 115408 (2013); V. Szaszk-Bogr, F. M. Peeters, P. Fldi, Phys. Rev. B
{\bf 91}, 235311 (2015).

\bibitem{RSOC2}  K. Hasanirokh, J. Azizi, A. Phirouznia, H. Mohammadpour, Eur. Phys. J. B {\bf 87}, 95 (2014).

\bibitem{cite:Han14}
W. Han, Roland K. Kawakami, M. Gmitra and J. Fabian, Nat. Nanotech., {\bf 9}, 794 (2014).

\bibitem{cite:Avsar14}
A. Avsar, et. al. Nat. Comm., {\bf 5}, 4875 (2014).

\bibitem{Kane05}
C. L. Kane and E. J. Mele, Phys. Rev. Lett. {\bf 95}, 226801 (2005).

\bibitem{so_exp1}
D. Marchenko, A. Varykhalov, M. R. Scholz, G. Bihlmayer, E. I. Rashba, A. Rybkin, A. M. Shikin, and O. Rader,
Nature Commun. {\bf 3}, 1232 (2012).

\bibitem{cite:Dyrdal}
A. Dyrdal, J. Barna\'{s}, and V. K. Dugaev, Phys. Rev. B  {\bf 89}, 075422 (2014).

\bibitem{cite:Bercioux}
D. Bercioux, F. D. Urban, F. Romeo, and R. Citro, Appl. Phys. Lett.  {\bf 101}
122405 (2012). 

\bibitem{cite:Gurjic}
M. M. Gruji\'{c}, M. Z. Tadi\'{c} and F. M. Peeters,
Phys. Rev. Lett. {\bf 113}, 046601 (2014).

\bibitem{RSOC4}
Y. Yao, F. Ye, X. Qi, S. Zhang, and Z. Fang, Phys. Rev. B {\bf 75}, 041401(R) (2007).

\bibitem{Zhai}
F. Zhai, Y. Ma and K. Chang, New J. Phys. {\bf 13} 083029 (2011).

\bibitem{cite:Makhlin01}
Yu. Makhlin, and A. D. Mirlin, Phys. Rev. Lett. {\bf 87} 276803 (2001); F. Zhou, B. Spivak and B. Altshuler,
Phys. Rev. Lett. {\bf 82} 608 (1999).

\bibitem{cite:Brouwer98}
P. W. Brouwer, Phys. Rev. B {\bf 58}, R10135 (1998); M. Moskalets,
M. Buttiker, Phys. Rev. B {\bf 66}, 035306 (2002).

\bibitem{cite:Pradaprb09} E. Prada, P. San-Jose and H.
Schomerus,  Phys. Rev. B  {\bf 80}, 245414 (2009); E. Prada, P. SanJose and H. Schomerus, Solid State Commun
{\bf 151}, 1065 (2011).

\bibitem{cite:Grichuk13}
E. Grichuk and E. Manykin,  Eur. Phys. J. B  {\bf 86}, 210
(2013).

\bibitem{Benjamin} C. Benjamin, Appl. Phys. Lett. {\bf 103}, 043120 (2013).

\bibitem{cite:Zhu09}
R. Zhu and H. Chen Appl. Phys. Lett.  {\bf 95} 122111 (2009).

\bibitem{cite:Low12}
T. Low, Y. Jiang, M. Katsnelson and F. Guinea, Nano Lett.
{\bf 12}, 850 (2012).

\bibitem{cite:abdollahipour14}
B. Abdollahipour and R. Mohammadkhani, J. Phys.: Condens. Matter,
{\bf 26}, 085304 (2014).

\bibitem{pmp1}
Q. Zhang, K. S. Chan, Z. Lin,  Appl. Phys. Lett. {\bf 98}, 032106 (2011);
Q. Zhang, K. S. Chan, Z. Lin, J. Liu, Phys. Lett. A {\bf 377}, 632 (2013);
J. Wang, K. S. Chan, and Z. Lin, Appl. Phys. Lett. {\bf 104}, 013105 (2014).

\bibitem{pmp2}
A. Kundu, S. Rao, and A. Saha, Phys. Rev. B {\bf 82}, 155441 (2010).

\bibitem{pmp3}
S. Singh , A. K. Patra , B. Barin , E. del Barco , and B. Ozyilmaz, EEE Trans. Mag. {\bf 49}, 3147 (2013).

\bibitem{so_exp2}
J. Balakrishnan, G. K. W. Koon, M. Jaiswal, A. H. Castro Neto and B. Ozyilmaz, Nat. Phys. 9, 284287 (2013).

\bibitem{cite:Pereira}
V. M. Pereira, A. H. Castro Neto and N. M. R. Peres, Phys. Rev. B  {\bf 80}, 045401 (2009).

\bibitem{cite:choi1}
S.-M. Choi, S.-H. Jhi and Y.-W. Son, Phys. Rev. B  {\bf 81}, 081407(R) (2010).

\bibitem{cite:soodchomshom}
B. Soodchomshom, Physica B {\bf 406}, 614 (2011);
B. Soodchomshom, J. Supercond. Nov. Magn. 24, 1715 (2011).

\bibitem{liuprb2007}
F. Liu, P. Ming, and J. Li, Phys. Rev. B {\bf 76}, 064120 (2007).

\bibitem{Lee08}
C. Lee, X. Wei, J. W. Kysar, J. Hone, Science 321, 385 (2008).

\bibitem{cite:Titov07}
M. Titov, Europhys. Lett. {\bf 79}, 17004 (2007).

\bibitem{aldst_prb_strain}
M. Alidoust, and J. Linder, Phys. Rev. B  {\bf 84}, 035407 (2011).

\bibitem{Covaci} L. Covaci and F. M. Peeters, Phys. Rev. B {\bf 84}, 241401(R) (2011).

\end{thebibliography}
\end{document}